\documentstyle[12pt]{article}
\newcommand{\zp}[3]{Z. Phys.\ C#1 (19#2) #3}
\newcommand{\pl}[3]{Phys.\ Lett.\ #1B (19#2) #3}
\newcommand{\np}[3]{Nucl.\ Phys.\ B#1 (19#2) #3}
\newcommand{\prd}[3]{Phys.\ Rev.\ D#1 (19#2) #3}
\newcommand{\prl}[3]{Phys.\ Rev.\ Lett.\ #1 (19#2) #3}

\newcommand{\md}{\mbox{d}}
\def\simgt{\rlap{\lower 3.5 pt \hbox{$\mathchar \sim$}} \raise 1pt \hbox {$>$}}
\def\simlt{\rlap{\lower 3.5 pt \hbox{$\mathchar \sim$}} \raise 1pt \hbox {$<$}}

\newcommand{\beq}{\begin{equation}}
\newcommand{\eeq}{\end{equation}}
\newcommand{\bea}{\begin{eqnarray}}
\newcommand{\eea}{\end{eqnarray}}

\catcode`\@=11

\def\section{\@startsection{section}{1}{\z@}{3.5ex plus 1ex minus .2ex}
{2.3ex plus .2ex}{\large\bf}}

\def\thesection{\arabic{section}.}

\def\appendix{\setcounter{section}{0}
 \def\thesection{Appendix \Alph{section}:}
 \def\theequation{\Alph{section}.\arabic{equation}}}
\def\figitem{\@ifnextchar[{\@lfigitem}{\@figitem}}
\def\@lfigitem[#1]#2{\item{Fig.~\@figlabel{#1}.}\if@filesw
      { \def\protect##1{\string ##1\space}\immediate
        \write\@auxout{\string\figcite{#2}{#1}}\fi\ignorespaces}}
\def\@figitem#1{\item\if@filesw \immediate\write\@auxout
       {\string\figcite{#1}{\the\c@enumi}}\fi\ignorespaces}
\def\figcite#1#2{\global\@namedef{fb@#1}{#2}}
\let\citation\@gobble

\def\fcite{\@ifnextchar [{\@tempswatrue\@fcitex}{\@tempswafalse\@fcitex{[]}}}
\def\@fcitex#1#2{\if@filesw\immediate\write\@auxout{\string\citation{#2}}\fi
  \def\@fcitea{}\@fcite{\@for\@fciteb:=#2\do
    {\@fcitea\def\@fcitea{,}\@ifundefined
       {fb@\@fciteb}{{\bf ?}\@warning
       {Figure `\@fciteb' on page \thepage \space undefined}}%
\hbox{\csname fb@\@fciteb\endcsname}}}{#1}}
\let\figdata=\@gobble
\let\figliststyle=\@gobble
\def\figlist#1{\if@filesw\immediate\write\@auxout{\string\figdata{#1}}\fi
  \@input{\jobname.figs}}
\def\figliststyle#1{\if@filesw\immediate\write\@auxout
    {\string\figliststyle{#1}}\fi}

\def\@fcite#1#2{#1\if@tempswa , #2\fi}

\def\thefiglist#1{
 \par\section*{Figures \@mkboth{Figures}{Figures}}
\list
   {\m@th{Fig.\ \arabic{enumi}.\ \hfill}}
   {\settowidth\labelwidth{Fig.\ \m@th {#1}.\ }%
    \leftmargin\labelwidth
    \advance\leftmargin\labelsep
    \usecounter{enumi}}
    \def\newblock{\hskip .11em plus .33em minus -.07em}
    \sloppy
    \sfcode`\.=1000\relax}

\def\@citex[#1]#2{\if@filesw\immediate\write\@auxout{\string\citation{#2}}\fi
  \def\@citea{}\@cite{\@for\@citeb:=#2\do
    {\@citea\def\@citea{,\penalty\@m}\@ifundefined
       {b@\@citeb}{{\bf ?}\@warning
       {Citation `\@citeb' on page \thepage \space undefined}}%
\hbox{\csname b@\@citeb\endcsname}}}{#1}}

\def\citer{\@ifnextchar [{\@tempswatrue\@citexr}{\@tempswafalse\@citexr[]}}

\def\@citexr[#1]#2{\if@filesw\immediate\write\@auxout{\string\citation{#2}}\fi
  \def\@citea{}\@cite{\@for\@citeb:=#2\do
    {\@citea\def\@citea{--\penalty\@m}\@ifundefined
       {b@\@citeb}{{\bf ?}\@warning
       {Citation `\@citeb' on page \thepage \space undefined}}%
\hbox{\csname b@\@citeb\endcsname}}}{#1}}
\relax
\begin{document}

\thispagestyle{empty}

\hfill\vbox{\hbox{\bf DESY 94-207}
            \hbox{\bf MZ-TH/94-28}
            \hbox{\bf TTP 94-20}
            \hbox{\bf hep-ph/9411372}
            \hbox{November 1994}
                                }
\vspace{0.5in}
\begin{center}
\boldmath
{\large\bf INELASTIC $J/\psi$ PHOTOPRODUCTION$\,^*$} \\
\unboldmath
\vspace{0.5in}
M.~Kr\"amer$^{a\dagger}$, J.~Zunft$^b$,
J.~Steegborn$^c$, and P.~M.~Zerwas$^b$ \\
\vspace{0.5in}
$^a$Inst.~Physik, Johannes Gutenberg-Universit\"at, D-55099 Mainz FRG\\
$^b$Deutsches Elektronen-Synchrotron DESY, D-22603 Hamburg FRG\\
$^c$Inst.~Theor.~Teilchenphysik, Univ. Karlsruhe, D-76128 Karlsruhe FRG\\

\end{center}
\vspace{0.5in}

\begin{center}
ABSTRACT \\
\end{center}

Inelastic photoproduction of $J/\psi$ particles at high energies
is one of the processes to determine the gluon distribution in the nucleon.
We have calculated
the QCD radiative corrections to the color-singlet model of this reaction.
They are large at moderate photon energies, but decrease with increasing
energies. The cross section and the $J/\psi$ energy spectrum are compared
with the available fixed-target photoproduction data and predictions are
given for the HERA energy range.

\vfill

\noindent
{\footnotesize
$*\,$ Supported in part by Deutsche Forschungsgemeinschaft DFG\\
$\dagger\,$ Present address: Deutsches Elektronen-Synchrotron DESY,
                             D-22603 Hamburg FRG;\\
\hspace*{3cm}E-mail: mkraemer@vxdesy.desy.de}
\newpage

\noindent
The measurement of the gluon distribution in the nucleon is one of the
important goals of lepton-nucleon scattering experiments. The classical
methods exploit the evolution of the nucleon structure functions with
the momentum transfer, and the size of the longitudinal structure function.
With rising energies,
however, jet physics and the production of heavy quark states become
important complementary tools. Besides open charm and bottom
production, the formation of $J/\psi$ bound states \cite{bj81}
in inelastic photoproduction experiments
\beq
\gamma + \cal{N} \to J/\psi + X
\eeq
provides an experimentally attractive method since $J/\psi$ particles are easy
to tag in the leptonic decay modes.

\vspace{3mm}

Many channels contribute to the generation of $J/\psi$ particles in
photoproduction experiments \cite{jst92}, similarly to hadroproduction
experiments. However, no satisfactory quantitative picture has emerged
yet and the production of a large surplus of $\psi'$ particles in
$p\overline{p}$ collisions awaits the proper understanding. Theoretical
interest so far has focussed on two mechanisms for $J/\psi$ photo- and
electroproduction, elastic/diffractive \cite{elastic,elastexp} and
inelastic production through photon-gluon-fusion \cite{bj81,jst92}.
While by the first mechanism one expects to shed light on the physical
nature of the pomeron,
inelastic $J/\psi$ production provides information on the distribution of
gluons in the nucleon \cite{mns87}. The two mechanisms can be separated
by measuring the $J/\psi$ energy spectrum, described by the scaling variable
\beq\label{eq_zdef}
z = \frac{pk_\psi}{pk_\gamma}
\eeq
with $p, k_{\psi,\gamma}$ being the momenta of the nucleon and $J/\psi$,
$\gamma$ particles, respectively. In the nucleon rest frame, $z$ is
the ratio of the $J/\psi$ to the $\gamma$ energy, $z=E_\psi/E_\gamma$. For
elastic/diffractive events $z$ is close to one; the inelastic events are
experimentally restricted in general to the range $z<0.9$ \cite{jung}.
The production of $J/\psi$ particles at large transverse momenta is dominated
by gluon fragmentation mechanisms \cite{bdfm94}.

Inelastic $J/\psi$ photoproduction through photon-gluon fusion is described in
the color-singlet model\footnote{Dual approaches to $J/\psi$ photoproduction
based on the subprocess $\gamma + g \to c + \overline{c}$ are, in contrast to
their $e^+e^-$ analogon, not sharply defined and they will therefore not be
discussed in the present context.} through the subprocess
\beq\label{eq_subpr}
\gamma + g \to J/\psi + g
\eeq
shown in Fig.\fcite{f_diagrams}.
Color conservation and the Landau-Yang theorem require
the emission of a gluon in the final state. The cross section is generally
calculated in the static approximation in which the motion of the charm quarks
in the bound state is neglected. In this approximation the production amplitude
factorizes into the short distance amplitude $\gamma + g \to c\overline{c}
+ g$, with $c\overline{c}$ in the color-singlet state and zero relative
velocity of the quarks, and
the $c\overline{c}$ wave function $\varphi(0)$ of the $J/\psi$ bound
state at the origin which is related to the leptonic width. The cross
section of the subprocess (\ref{eq_subpr}) may be written as
\beq\label{eq_jpsigb4dim}
\frac{\md\sigma^{(0)}}{\md t_1} =
\frac{128\pi^2}{3}\, \frac{\alpha\alpha^2_s e_c^2}{s^2}\,
 M^{2}_{J/\psi}\, \frac{|\varphi(0)|^2}{M_{J/\psi}}\;
\frac{s^2\,s_1^2 + t^2\,t_1^2 + u^2\,u_1^2}{s_1^2\;t_1^2\;u_1^2}
\eeq
where $t$ denotes the momentum transfer squared from the photon to the $J/\psi$
particle, and $s$,$u$ are the total energy and the momentum transfer from the
photon to the gluon [$t_1 = t - M_{J/\psi}^{2}$ {\em etc.}]. The cross section
(\ref{eq_jpsigb4dim})
is infrared finite. The photoproduction cross section on the nucleon
is obtained by integrating the subprocess over the gluon flux,
\beq
\sigma = \int\md\xi\, g(\xi)\,\hat\sigma
\eeq
When confronted with photoproduction data of fixed-target experiments
\cite{na14,ftps},
the theoretical predictions underestimate the measured cross section in general
by more than a factor two, depending in detail on the $J/\psi$ energy and the
choice of the parameters [for a recent review see Ref.~\cite{jst92}].
The discrepancy with cross
sections extrapolated from electroproduction data \cite{emc,nmc}
is even larger.

The lowest-order approach to the color-singlet model demands several
theoretical refinements: (i) Higher-order perturbative QCD corrections; (ii)
Relativistic corrections due to the motion of the charm quarks in the $J/\psi$
bound state; and last but not least, (iii) Higher-twist effects which are not
strongly suppressed due to the fairly low charm-quark mass. While
the relativistic corrections have recently been demonstrated to be under
control
in the inelastic region \cite{jkgw93},
the problem of higher-twist contributions has
not been approached so far. The analysis of the higher-order perturbative
QCD corrections will be presented in this note. Expected {\em a priori} and
verified subsequently, these corrections by far dominate the relativistic
corrections in the inelastic region, being of the order of several
$\alpha_s(M_{J/\psi}^{2})\sim 0.3$. In the first
step of a systematic expansion,
they can therefore be determined in the static approach \cite{bbl94}.

Generic diagrams which build up the cross section in next-to-leading order
are depicted in Fig.\fcite{f_diagrams}.
Besides the usual self-energy diagrams and
vertex corrections for photon and gluons (b), we encounter box diagrams (c),
the splitting of the final-state gluon into gluon and light quark-antiquark
pairs, as well as diagrams renormalizing the initial-state parton densities
(e). The evaluation of these amplitudes has been performed in the
Feynman gauge.
We have adopted the dimensional regularization scheme to calculate the singular
parts of the amplitudes. [We have
refrained from using the seemingly simpler scheme
of dimensional reduction which however gives rise to complications in the
massive quark case as pointed out in Ref.~\cite{bkns89}.]
The masses of light quarks
in Fig.\fcite{f_diagrams}(d,e$_1$) have been neglected while the mass
parameter of the charm
quark has been defined on-shell, with $m_c=\frac12 M_{J/\psi}$ for
consistency. Since the proper renormalization scale $\mu_R$ is expected in the
range of about $m_c$ to $M_{J/\psi}= 2 m_c$, we have used the
$\overline{\mbox{\em MS}\hphantom{.}}$ {\em scheme}
\cite{msbar} with four active flavors for the
renormalization of the QCD coupling $\alpha_{s}^{(n_f)}(\mu_{R}^{2})$.
We have
carried out the renormalization program also in the {\em extended}
$\overline{\mbox{\em MS}\hphantom{.}}$ {\em scheme} \cite{msbarext}
in which the massive particles are decoupled smoothly for momenta smaller
than the quark mass.
The results
in the two schemes differ marginally, by typically less than 6 \%.
The exchange of Coulombic gluon quanta in the diagram (1c) leads to a Coulomb
singularity $\sim \pi^2/2\beta_R$
which can be isolated by introducing a small relative quark velocity
$\beta_R$.
Following the standard path \cite{hb57},
we interpret this effect
as Sommerfeld rescattering correction which can effectively be mapped into the
$c\overline{c}$ wave function. As expected, the infrared
singularities cancel
when the emission of soft and collinear final-state gluons and light quarks,
characterized by a cut-off $\Delta$ \cite{bkns89,sn92}, is added to the
virtual corrections.
The collinear initial-state singularities can be absorbed, as usual, into the
renormalization of the parton densities \cite{aem79} defined in the
$\overline{\mbox{MS}}$ factorization scheme.

After carrying out this straightforward but tedious program, the results for
the cross sections of the subprocesses can be presented in the form of
scaling functions,
\beq
\hat\sigma_{i\gamma}(s,m_c^2) =
\frac{\alpha\alpha_s^2 e^2_c}{m_c^2}\,\frac{|\varphi(0)|^2}{m_c^3}
\left[ c_{i\gamma}^{(0)}(\eta) + 4\pi\alpha_s \left\{c_{i\gamma}^{(1)}(\eta)
+\overline{c}^{(1)}_{i\gamma}(\eta)\ln\frac{Q^2}{m_c^2}\right\}\right]
\eeq
$i=g,q,\overline{q}$ denoting the parton targets. For the sake of simplicity,
we have identified the renormalization scale with the factorization scale
$\mu_R^2 = \mu_F^2 = Q^2$. The scaling functions depend
on the energy variable $\eta = s/4m_c^2 - 1$. $c_{\gamma g}^{(0)}$ is the
lowest-order
contribution which scales $\sim \eta^{-1}\sim 4m_c^2/s$ asymptotically.
$c_{\gamma g}^{(1)}$ can be decomposed into a "virtual + soft" (V+S) piece,
and a "hard" (H) gluon-radiation piece; the $\ln^j\Delta$ singularities of the
(V+S) cross section are mapped into (H), cancelling the equivalent
logarithms in this contribution so that the limit $\Delta\to 0$ can safely be
carried out \cite{sn92}.
The {\it nomenclatura} "hard" and "virtual + soft" is therefore a
matter of definition, and negative values of $c^{(\mbox{\scriptsize{}H})}$
may occur in some regions of the parameter space. Up to this order, the
wave-function at the origin is related to the leptonic $J/\psi$ width by
\beq\label{eq_decaycorr}
\Gamma_{ee} = \left(1 - \frac{16}{3}\frac{\alpha_s}{\pi}\right)
\frac{16\pi\alpha^2e_c^2}{M_{J/\psi}^{2}}\, |\varphi(0)|^2
\eeq
with only transverse gluon corrections taken into account explicitly
\cite{lepdecay}.

The scaling functions $c_{\gamma i}(\eta)$ are shown in
Figs.\fcite{f_scale}a/b for
the parton cross sections integrated over $z\le z_1$ where we have chosen $z_1
= 0.9$ as discussed before. [Note that the definition of $z$ is the same at the
nucleon and parton level since the momentum fraction $\xi$ of the partons
cancels in the ratio (\ref{eq_zdef}).] The following comments can be inferred
from the figures. (i) The form of the scaling functions resembles the scaling
functions
in open-charm photoproduction \cite{sn92}. However, there is an important
difference. The "virtual + soft" contribution for $J/\psi$ production is
significantly more negative than for open-charm production. The destructive
interference with the lowest-order amplitude is not unplausible though,
as the momentum transfer of virtual gluons has a larger
chance [in a quasi-classical approach] to scatter quarks out of the small
phase-space element centered at $p_c +
p_{\overline{c}} = p_{J/\psi}$ than to scatter them from outside into this
small element.
(ii) While $c_{\gamma g}^{(0)}$ and
$c_{\gamma g}^{(\mbox{\scriptsize{}1,V+S})}$ scale asymptotically $\sim 1/s$,
the hard coefficients
$c_{\gamma g}^{(\mbox{\scriptsize{}1,H})}$ and
$c_{\gamma q}^{(1)}$ [as well as $\overline{c}_{\gamma g}^{(1)}$] approach
plateaus for high energies, built-up by the flavor excitation mechanism.
(iii) The cross sections on the quark targets are more than one order of
magnitude
smaller than those on the gluon target. (iv) A more detailed presentation
of the spectra would reveal
that the perturbative analysis is not under proper control in the limit $z\to
1$, as anticipated for this singular boundary region \cite{phd}.
Outside the diffractive
region, i.e. in the truly inelastic domain, the perturbation theory is
well-behaved however.

The cross sections for $J/\psi$ photoproduction on nucleons are presented in
Figs.\fcite{f_zdist}a and \fcite{f_zdist}b.
In the first of the two figures we compare the
leading-order and next-to-leading order calculations with the $J/\psi$ energy
spectra of the two fixed-target photoproduction experiments at photon energies
near $E_\gamma = 100$~GeV, corresponding to invariant energies of about
$\sqrt{s\hphantom{tk}}\!\!\!\!\! _{\gamma p} \; \approx\; 14$~GeV. The GRV
parametrizations of the parton densities \cite{grvprot}
have been used. Since the
average momentum fraction of the partons $< \xi > \sim 0.1$ is moderate, the
curves are not sensitive to the parametrization in the small-$x$ region. The
results are shown for two values of
$\alpha_s^{(4)}(M_{J/\psi}^{2}) = 0.25$ and $0.30$ which correspond to the
average fit value in Ref.\cite{altac} and
the $1\sigma$ upper boundary of the error band,
respectively. Since the cross section depends strongly on the QCD coupling, we
adopt this measured value,
thus allowing for a slight inconsistency to the extent that
the GRV fits are based on a marginally lower value of $\alpha_s$. The
$K$-factor
$K=\sigma_{\mbox{\scriptsize{}NLO}}/\sigma_{\mbox{\scriptsize{}LO}}$
turns out to be $\sim 2.45$ with one part $\sim$ 1.73 due to the QCD
radiative corrections of the leptonic $J/\psi$ width and a second part
$\sim$ 1.44 due to the
dynamical QCD corrections {\em per se}. The $K$-factor is nearly independent of
$z$. The dependence on the renormalization/factorization scale $Q$ is reduced
considerably in next-to-leading order \cite{phd}. While the ratio of the cross
sections in leading order for
$Q=m_c : (\sqrt{2}\,m_c) : M_{J/\psi}$
is given by
$1.7 : 1.3 : 1$, it is much closer to unity, $0.9 : 1.1 : 1$,
in the next-to-leading order calculation, Fig.\fcite{f_qsqdep}.
The overall behavior of the $Q$ dependence is reminiscent of heavy quark
production in hadron collisions \cite{admn88}.
The cross section runs through a maximum \cite{pms} near $Q\approx 1.4\;m_c$
with broad width, the origin of the stable behavior in $Q$.
In the BLM scheme \cite{blm} $Q$ moves from values below $m_c$ at
low energies up to $\sim\, 1.5 \; m_c$ at the \mbox{HERA} energy of
$\sqrt{s\hphantom{tk}}\!\!\!\!\! _{\gamma p} \; \approx\; 100$~GeV.
In particular the value at high energies is significantly
larger than the
corresponding BLM value for $J/\psi$ decays. The typical kinematical
energy scale is not set any more by the small gluon energy in the $J/\psi$
decay but rather by the typical initial-state parton energies.

In a systematic expansion we may finally add the relativistic corrections as
estimated in Ref.\cite{jkgw93}.
Two conclusions can be drawn from the final results presented
in Fig.\fcite{f_zdist}a.
(i) The $J/\psi$ energy dependence
d$\sigma/\mbox{d}z(\gamma + \cal{N}
\to J/\psi + X)$ is adequately accounted for
by the color-singlet model so that the shape of the gluon distribution in the
nucleon can be extracted from $J/\psi$ photoproduction data with confidence.
(ii) The absolute normalization  of the cross section is somewhat less certain;
this is apparent from the comparison with the photoproduction data. [The
situation is worse for electroproduction data
\cite{jst92}].
However, allowing for higher-twist uncertainties of order $(\Lambda/m_c)^k
\;\simlt\; 20\%$ for $k \ge 1$, we conclude that the normalization too appears
to be under semi-quantitative control -- at the least.

In Fig.\fcite{f_zdist}b we present the prediction of the cross section for the
\mbox{HERA} energy range. In this high energy range the $K$-factor is smaller
than at low energies, a consequence of the negative dip in the
$c^{(1)}$ scaling function of Fig.\fcite{f_scale}. For
$\alpha_s^{(4)}(M_{J/\psi}^{2}) = 0.25$ we find, for $z < 0.9$,
a value of about
$\sigma(\gamma + p \to J/\psi + X) \approx 21$~nb at an invariant
$\gamma p$ energy of $\sqrt{s\hphantom{tk}}\!\!\!\!\! _{\gamma p}
\; \approx\; 100$~GeV; this value rises to about 32~nb if we choose the
larger value 0.30 for the QCD coupling. The production of $\Upsilon$ bottonium
bound states is suppressed by a factor of about 300 at \mbox{HERA}, a
consequence of the smaller bottom electric charge and the phase space
reduction by the large $b$ mass.

\nopagebreak
{\em In summa.} We have shown in this next-to-leading order perturbative QCD
analysis that the energy shape of the cross section for $J/\psi$
photoproduction is
adequately described by the color-singlet model. A semi-quantitative
understanding, at the least,
has been achieved for the absolute normalization of
the cross section. Higher-twist effects must be included to improve the
quality of the theoretical analysis further. The predictions for the
\mbox{HERA} energy range provide a crucial test for the underlying picture as
developed so far in the perturbative QCD sector.

\vspace{2cm}

{\noindent\large\bf Acknowledgements}
\vspace{3mm}

\noindent
We have benefitted from discussions with S.J.\ Brodsky, J.H.\ K\"uhn,
G.\ Schuler and W.\
van Neerven. Special thanks go to W.\ Beenakker, V.\ Del Duca and M.\ Spira
for their continuous advice in solving $\epsilon$ problems. Last but not least,
discussions on photo- and electroproduction data with H.\ Jung and J.N.\ Lim
are gratefully acknowledged.
MK would like to thank J.G.\ K\"orner for discussions and support.

\vspace{3mm}



%

\begin{thefiglist}{99}
\figitem{f_diagrams}
Generic diagrams for inelastic $J/\psi$ photoproduction:
(a) leading order contribution;
(b) vertex corrections;
(c) box diagrams;
(d) splitting of the final state gluon into gluon or
    light quark-antiquark pairs;
(e) diagrams renormalizing the initial-state parton densities.
\figitem{f_scale}
(a)
Coefficients of the QCD corrected total inelastic [$z < 0.9$] cross section
$\gamma + g \to  J/\psi + X$ in the physically relevant range of the scaling
variable $\eta = s_{\gamma p}/4m^2 - 1$; and (b) for
$\gamma + q/\overline{q} \to  J/\psi + X$.
\figitem{f_zdist}
(a) Energy spectrum $\md\sigma/\md{}z$,
at the initial photon energy $E_\gamma =
100$~GeV, for two different values of $\alpha_s(M^{2}_{J/\psi})$ compared
with the photoproduction data \cite{na14,ftps}.
(b) Total cross section for inelastic $J/\psi$ photoproduction
    $\gamma + P \to  J/\psi + X$ as a function of the
    photon-proton center of mass energy in the HERA energy range.
    [The dashed line presents the $\Upsilon$ photoproduction cross
    section, amplified by a factor 100.]
\figitem{f_qsqdep}
Dependence of the total cross section $\gamma + g \to  J/\psi + X$
on the renormalization/factorization scale $Q$
at a photon energy of $E_\gamma = 100$~GeV.
\end{thefiglist}

\end{document}